# Threat-Agnostic Resilience: Framing and Application for Critical Infrastructure

Benjamin D. Trump[1], Stergios Mitoulis[2], Sotirios Argyroudis[3], Gregory Kiker[4], José Palma-Oliveira[5], Robert Horton[4,6], Gianluca Pescaroli[7], Elizaveta Pinigina[8], Joshua Trump9, Igor Linkov[4, 10*]

1 University of Michigan

2 University of Birmingham

3 Brunel University

4 University of Florida

5  University of Lisbon

6 Dallas-Fort Worth International Airport

7 University College London

8 London School of Economics

9 Virginia Tech

10  Carnegie Mellon University

*corresponding author, ilinkov@yahoo.com

## Abstract
Critical infrastructure is not indestructible. Threats, both emergent and systemic, have propagated beyond historical norms that risk mitigation efforts alone cannot alleviate. Interdependencies between infrastructural systems compound consequences at vulnerable locations but can be harnessed to maximize operational efficiency and recovery capability. Characterizing practical forms of resilient infrastructure through 5 basic principles – modularity, distributedness, redundancy, diversity, and plasticity – provides a foundation for critical infrastructure systems to absorb, recover, and adapt to disruptions agnostic of threat source. Challenges exist in developing methodological foundations for these principles within practical applications to prevent sunk cost and over-constraining operational procedures. This study evaluates each principle, establishing practical forms, quantification strategies, and their impact on critical infrastructure resilience.

## Introduction
In an era marked by rapid technological advancements and increasing interconnectivity, the resilience of critical infrastructure has become a paramount concern[1,2]. Critical infrastructure encompasses the systems and assets essential to the functioning of societies, including power grids, transportation networks, communication systems, and water supply networks. The Cybersecurity and Infrastructure Security Agency (CISA) explores infrastructure's role within national critical functions that enable economic stability, public safety, and national security[3]. Disruption to infrastructure can lead to cascading failures with profound socio-economic impacts on national critical functions[4,5]. The landscape of possible threat to infrastructure and critical functions has broadened, ranging from natural disasters and cyber-attacks to human errors and equipment failures[6,7]. Despite growing, unpredictable threats, infrastructure must be resilient to prevent national disruption such as the Texas Freeze.

Resilience, in the context of critical infrastructure, refers to the ability of infrastructure to prepare for threats and hazards, adapt to changing conditions, and recover from disruptions. Traditional approaches to resilience have often focused on threat-specific strategies, preparing for and mitigating the impacts of identified hazards and stressors[8,9]. While this method has its merits, it falls short in addressing the unpredictability and diversity of contemporary threats. A more holistic approach is needed—one that encompasses resilience to both known and unknown threats. This is where the concept of threat-agnostic resilience emerges as a critical paradigm.

Threat-agnostic resilience refers to the ability of a system to maintain its critical functions regardless of the specific nature of the threat it faces. This approach transcends the limitations of threat-specific risk assessment and scenario-based resilience evaluation by focusing on the inherent qualities and capabilities of the system itself. By emphasizing system characteristics that contribute to overall robustness and adaptability, threat-agnostic resilience aims to ensure continuous operation of a system's critical functions in the face of unforeseen challenges and/or disruptions to complex, interdependent systems[10].

Threat-agnostic resilience is not a static property, but rather a dynamic and evolving capability[11]. As the nature and intensity of threats continue to change over time, so too must the characteristics and network properties of a system's resilience. This type of assessment is an iterative process and must involve stress-testing to optimize and balance the characteristics of a system's resilience[12], such as: instituting modular system connections, distributing system resources to prevent nodal collapse, implementing redundant architectures for backup planning, diversifying agents, and incorporating adaptive response within agents as system plasticity.

The primary objective of this article is to elucidate the characteristics that inform threat-agnostic resilience in critical infrastructure. By identifying and analyzing these characteristics, we aim to provide a framework for assessing and enhancing the resilience of complex systems. The framework presented in this article should be viewed as a starting point for guiding the identification and cultivation of threat-agnostic resilience characteristics, rather than a prescriptive, one-size-fits-all blueprint. By tailoring and adapting this framework to the specific needs and circumstances of each infrastructure system, we can develop customized resilience strategies that are effective, efficient, and sustainable over the long term[13].

## Characteristics of Threat-Agnostic System Resilience

The development of threat-agnostic resilience in critical infrastructure systems relies on the identification and cultivation of specific system characteristics. The key characteristics that have emerged include modularity, distributedness, redundancy, diversity, and plasticity. These characteristics, when properly integrated into the design and operation of infrastructure systems, contribute to their ability to maintain functionality and integrity in the face of diverse and unpredictable threats, such as environmental, cyber, anthropogenic, and geopolitical conflicts[12,14,15]. These disruptions cause schisms within infrastructural integrity at various domains, including within physical infrastructure, social response, cyber components, and financial health[16]. The resilience characteristics proposed in this research contribute to maintaining the integrity of infrastructure across each domain despite the unpredictability of threat origin, i.e., threat agnosticity (Figure 1).

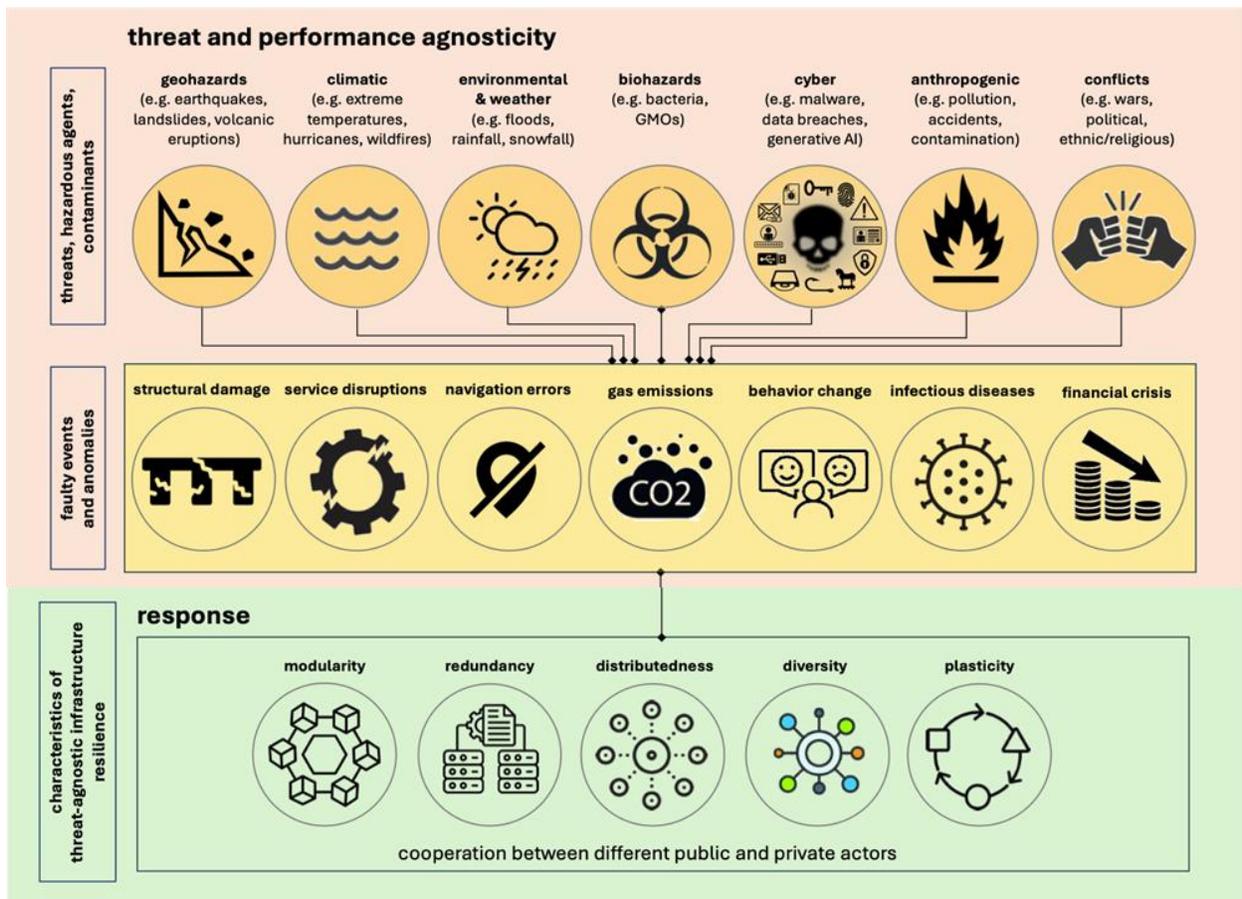

**Figure 1**. Framework for assessing and enhancing threat-agnostic resilience of complex systems. Unpredictable threats (orange circles) have a variety of impacts across domains of critical infrastructure (yellow circles). Implementing resilience across domains requires threat-agnostic resilience characteristics (green circles).

Table 1 presents an overview of each principle of threat-agnostic resilience. The following sections dive into greater detail the components of each principle's definition, practical forms of each principle in critical infrastructure, quantification strategies in various fields, and the contribution to threat-agnostic resilience.

Table 1. Characteristics of threat-agnostic resilience

| Threat-agnostic resilience characteristic | | | |
|---|---|---|---|
| Definition | Practical examples | Quantification strategies | Contribution to threat agnostic resilience |
| Modularity | | | |
| Degree to which system components can be separated, reengineered and recombined. | Modular Design, Construction, and Repair | Modularity and coupling indices, clustering coefficient, average path length | Limits impact of localized failures; facilitates easy replacement and upgrading. Each module is often considered 'plug and play', or easily substituted with minimal startup resources. |
| Distributedness | | | |
| Distribution of system functions across multiple nodes or components, reducing reliance on a central service or authority. | Decentralized functions and control, load balancing | Centrality, clustering coefficient, average path length, diameter | Eliminates single points of failure and enabling local services and responses to maintain system functionality. |
| Redundancy | | | |
| Duplication of critical components or functions to increase capacity and reliability with parallel components or functions (not in series). | Redundant architectures, "n" redundancy, network redundancy | Redundancy ratio, connectivity index, | Ensures availability of alternative resources regardless of threat type; provides multiple layers of service and protection. |
| Diversity | | | |
| Inclusion of diverse components or strategies to handle a variety of threats and consequences. | Heterogeneous Systems | Shannon index, functional diversity index, qualitative indicators | Increases the likelihood of some components surviving or functioning under different threat conditions. |
| Plasticity | | | |
| Ability of a system to adapt its structure or behavior in response to changes in the environment or internal conditions. | Dynamic Reconfiguration; system upgrade; versatility | Adaptability index, reconfigurability index, qualitative indicators | Enables dynamic response to unforeseen threats and supports continuous operation by reconfiguring component behavior, resources and strategies. |

## Modularity

Modularity is a key principle in the design of resilient engineered systems. It allows for the decomposition of complex systems into smaller, more manageable components. Each module can be designed, developed, and tested independently, while still maintaining the ability to integrate seamlessly with other modules to form a cohesive, integral system. This approach enhances the overall resilience of the system by localizing potential failures and enabling rapid recovery through the replacement or repair of individual modules without affecting the entire system and its operations. It is critical to note that modularity, as well as any resilience principle, alone cannot guarantee the resilience of an engineered system to all threats. Specifically, modularity alone may fail to recover from systemic disruption within highly interconnected environments, e.g., common problems such as border conflicts surrounding water scarcity.

In practice, modularity can be achieved through the application of standardized interfaces, protocols, and architectures. These standards ensure interoperability between different modules and facilitate the plug-and-play integration of components from various vendors. For instance, in a modular water distribution system, standardized pipe fittings and valve configurations allow for the easy connection and disconnection of different subsystems, such as treatment plants, storage tanks, and distribution networks. This modularity enables the system to adapt to changing demands and maintain functionality even when individual components fail.

To quantify the modularity of an infrastructure system, network science provides a range of metrics and tools such as the modularity index, coupling index, clustering coefficients, and average path length. The modularity index measures the strength of division of a network into modules or communities, where higher scores indicate connected modules with sparse inter-module connections, which is a desirable property for resilient systems. The coupling index quantifies the degree of interdependence between modules[17], where lower scores suggest modules have limited co-dependence. Clustering coefficients and average path lengths in tandem describe how tightly-knit infrastructure systems are. A high clustering coefficient indicates the presence of functional modules, while a low path length represents efficient resource flow between modules.

Two benefits of modularity include the ability to scale the system over time and to decouple functions within the infrastructure system. Scalability is particularly important in rapidly growing urban areas, where the infrastructure needs to keep pace with the increasing population, economic activity and changing demand. For example, in a modular transportation system, new bus routes or train lines can be added to the network without disrupting the existing services, thereby enhancing the overall capacity and resilience of the system. Decoupling functions expands system scaling by optimizing parts of the whole, for example, modular power distribution systems can separately optimize generation, transmission, and distribution systems without requiring complete system overhauls.

## Distributedness

Where modularity refers to the capacity of engineered systems to operate in discrete, self-contained compartments, distributedness refers to the allocation of system functions and governance across multiple dispersed nodes or components. This design reduces reliance on a central authority or single point of control, enhancing the system's ability to operate independently in different locations. Distributedness also enables the scalability of infrastructure systems, where new nodes or components can be added to the network without requiring significant modifications to the existing architecture. This capability allows for the gradual expansion and upgrading of the system over time, in response to changing demands or technological advancements.

In the context of practical infrastructure systems, distributedness can be achieved through the implementation of distributed control architectures, such as multi-agent systems[18], peer-to-peer networks, or blockchain-based platforms. Distributed optimization algorithms, such as consensus-based methods or alternating direction method of multipliers (ADMM), can be applied to achieve system-wide objectives, such as energy efficiency or load balancing[19]. Within a fully distributed infrastructure system, each node or component has the capacity to process information, make decisions, and coordinate actions with other nodes in the network. For example, in a distributed water management system or transport system, smart sensors and actuators can be deployed throughout the network to monitor service levels, such as water quality or traffic conditions in real-time, without relying on a central control room.

To quantify the degree of distributedness in an infrastructure system, various metrics from network science can be applied. One commonly used metric is the degree of centrality, which measures the extent to which the functionality of the system depends on a few central nodes. A lower degree of centrality indicates a more distributed system, where the importance and influence of individual nodes are more evenly spread across the network. Similar to modularity, the clustering coefficient and average path length provide location and commodity-based views of infrastructure distribution. Furthermore, network diameter represents the maximum distance between any pair of nodes. A distributed system with a low average path length and a small diameter can facilitate the rapid dissemination of information and the efficient coordination of actions across the network, even in the presence of failures or disruptions.

One of the key advantages of distributedness in infrastructure systems is the increased resilience to failures and attacks. In a centralized system, a failure or compromise of the central node can lead to the collapse of the entire system. In contrast, a distributed system can continue to function even if some of its nodes are damaged or disconnected, as the remaining nodes can compensate for the loss in service and control and maintain the system's overall functionality at acceptable levels. This resilience is particularly important in the face of natural disasters, cyber-attacks, or other disruptions that can target specific components of the infrastructure.

## Redundancy

Redundancy ensures the availability of backup components or functions in the event of failures or disruptions[20]. For engineered infrastructure, redundancy can be achieved through the duplication or replication of critical assets, such as power generators, communication links, or water treatment plants. This redundancy allows the system to maintain its functionality even when some components fail, by seamlessly switching to the backup components or rerouting the flow of resources through alternative paths of similar constitution or service delivery.

The implementation of redundancy in infrastructure systems can take various forms, depending on the specific requirements and constraints of the system. One common approach is the use of N+1 or N+2 redundancy, where N represents the number of components required for normal operation, and the additional components serve as backups[21]. For example, in a data center with N+1 redundancy, if one server fails, the backup server can immediately take over its functions without interrupting the services provided by the data center. Redundancy can also be achieved at the system level, by providing multiple alternative paths or routes for the flow of resources, such as electricity, transportation of people and goods, water, or data. This type of redundancy, known as network redundancy or path diversity, enhances the resilience of the system to link failures or congestion[22]. In a transportation network, for instance, the presence of multiple alternative routes between origin and destination points allows for the rerouting of traffic in the event of road closures or accidents. Similarly, in a communication network, the deployment of redundant fiber optic cables or wireless links ensures the continuity of data transmission, even if some links are damaged or degraded.

To quantify the level of redundancy in an infrastructure system, several metrics can be employed. One widely used metric is the redundancy ratio, which is defined as the ratio of the number of redundant components to the total number of components in the system[23]. A higher redundancy ratio indicates a more redundant system, which is generally more resilient to failures. For example, a power grid with a redundancy ratio of 0.2 means that 20% of its components are redundant, providing a significant buffer against potential failures. Another important metric for assessing the redundancy of an infrastructure system is the connectivity index, which measures the number of independent paths between any two nodes in the network. A higher connectivity index suggests a more redundant and resilient system, as it indicates the presence of multiple alternative routes for the flow of resources[24].

An advantage of redundancy to infrastructure resilience is the presence of physical back-ups as responsory action to disruption but is presented with opportunity cost. In general, higher levels of redundancy provide greater resilience, but also incur higher costs in terms of capital investment, maintenance, and operation[21,25]. Therefore, the design of redundant systems should involve a trade-off analysis between the benefits of increased resilience and the associated costs, taking into account the specific requirements and constraints of the system.

## Diversity

In the context of infrastructure systems, diversity refers to the incorporation of heterogeneous components, technologies, and operational strategies, which collectively enhance the system's ability to withstand disruptions and maintain its functionality under varying conditions[26]. Diversity is a critical characteristic of resilient engineered systems, as it enables them to cope with a wide range of threats and uncertainties[27].

Diversity can be achieved in infrastructure systems through varying source providers, incorporating heterogeneous components and materials, and assimilating operational strategies and control mechanisms. Varying source providers within power generation might include a diverse mix of renewable energy sources, such as solar, wind, and hydro, can be integrated alongside conventional generators, providing a hedge against market fluctuations and geopolitical risks[28]. Implementing various infrastructural materials or ways of throughput also increases resilience to single modes of failure, such as by incorporating heterogeneous pipe materials in water distribution networks[29,30] or adopting multiple transportation networks within smart cities. Diversifying control strategies[31], such as in distributed power generation from smart grids or adaptive signal control in transportation engineering, can minimize strain on the reference system during peak periods and allow for continuous flow of goods, people, or resources[32].

To quantify the level of diversity in an infrastructure system, metrics such as the Shannon index and functional diversity index are viable. The Shannon index measures the richness and evenness of different types of components in the system[33], and while the Shannon index is commonly used in ecology, it can be used within infrastructure to assess the impact of infrastructure improvements on environmental diversity[34]. A higher Shannon diversity index indicates a more diverse system, which is generally more resilient to threats and uncertainties. In parallel, the functional diversity index measures the variety and distribution of different functional attributes, such as the capacity, efficiency, or reliability of each component or subsystem[35]. A higher functional diversity index suggests a more versatile and adaptable system, which can maintain its performance under different conditions and requirements.

In addition to these quantitative metrics, the assessment of diversity in infrastructure systems should also consider the qualitative aspects of the system's resilience[36]. For example, the compatibility and interoperability of different components and technologies should be evaluated, to ensure that they can work together seamlessly and efficiently[37]. The scalability of the system should also be assessed, to determine its ability to accommodate future growth and adapt to changing demands.

An advantage of diversity within infrastructure systems is the increased likelihood of system survival and operation to any disruption. By embracing diversity, infrastructure systems can reduce their reliance on any single component or technology and improve their adaptability to changing environments. However, overly diverse systems can limit other resilience characteristics, such as modularity and plasticity, potentially making improvements to system capacity and quality more laborious and resource-intensive.

## Plasticity

Plasticity enables engineered systems to adapt their structure or behavior in response to changing conditions. While plasticity holds definitions in other fields such as materials science, in the context of infrastructure systems for this research, plasticity refers to the ability of the system to modify its configuration, operations, or performance based on the dynamic variations in the environment, user demands, or internal states without further degradation of the system's performance after a disruption. Other terms in network science and ecology that are adjacent to this definition include suppleness (the ability of a network to maintain form under stress[38]) and adaptability (actors' influence on resilience within a system[39]). However, the definition of plasticity posited by this paper combines the influence of both *actor* and *network* to impose systemwide change before, during, or after a disruption.

Practical applications of plasticity in infrastructure systems includes adaptive control strategies, reconfigurable architectures, and mechanisms for self-organization. Adaptive control involves real-time monitoring and adjustment of system parameters based on feedback loops and learning algorithms[40]. For example, in a smart energy grid, adaptive control can be used to dynamically balance the supply and demand of electricity, by optimizing the dispatch of generators, the configuration of transmission lines, or the pricing of energy services[41]. Reconfigurable architectures allow the system to change its structure or topology, by adding, removing, or rearranging its components or connections. For instance, in a modular transportation network of high plasticity, reconfigurable architectures can be used to dynamically adjust the layout of roads, bridges, or terminals, based on the shifting patterns of traffic flow, land use, or urban development[42]. Self-organizing mechanisms rely on the local interactions and autonomous behaviors of the system components, which collectively give rise to the emergence of global patterns and functions. For example, in a decentralized water distribution network, self-organizing mechanisms can be used to enable the autonomous coordination of pumps, valves, and tanks, based on the local sensing and communication of water quality, pressure, or demand.

To assess the plasticity of an infrastructure system, metrics, such as the adaptability index and reconfigurability index can be adopted. The adaptability index measures the degree to which the system can modify its structure or behavior in response to perturbations[43]. It is a function of the range and speed of the system's responses, as well as the effectiveness and efficiency of the adaptations. A higher adaptability index indicates a more plastic system, which can better cope with the changing conditions and maintain its performance over time. The reconfigurability index quantifies the ease and speed with which the system can be reconfigured to meet new requirements or recover from failures[44] and is a function of the number and diversity of the system's configurations, as well as the time and cost required for the reconfigurations. A higher reconfigurability index suggests a more flexible, responsive system, such as highway segments after a major flood.

In addition to these quantitative metrics, the assessment of plasticity in infrastructure systems should also consider the qualitative aspects of the system's resilience. For example, the robustness and scalability of the adaptive control strategies should be evaluated, to ensure that they can handle a wide range of perturbations and uncertainties, without leading to unintended consequences or cascading failures. The interoperability and compatibility of the reconfigurable architectures should also be assessed, to ensure that they can seamlessly integrate with the existing systems and standards, while enabling the smooth transition between different configurations.

Plastic infrastructure configurations are advantageous through resilience-by-design and resilience-by-intervention principles[45,46]. By incorporating system architectures that are innately adaptive, infrastructure layouts can inherently self-organize and can implement agents for response that presume multiple roles. However, a balancing point for consideration by practitioners might be the quality of adaptive architectures and the time-to-survive[47] during a recombination period for the uptake of new roles by plastic agents.

## A Framework for Adopting Resilience Characteristics within Critical Infrastructure Systems

A conceptual framework is required to implement threat-agnostic resilience into infrastructure systems. This process involves an assessment of how each resilience characteristic contributes to maintaining operation of critical functions under various perturbations. As defined by the Cybersecurity and Infrastructure Security Agency (CISA), critical functions are the functions of government and the private sector that are paramount to a nation's security, economic health, and public health and are commonly upheld by critical infrastructure[48]. Evaluating critical infrastructure in the broader service-level lens begins an assessment of interconnectivity within critical infrastructure across sectors, which is crucial quantifying resilience within a network.

To operationalize threat-agnostic resilience, we propose a multi-step framework that integrates network science and systems engineering principles (Figure 2). The first step involves a comprehensive analysis of the infrastructure system to identify its critical functions and the infrastructure that supports them. For instance, in a power grid, critical functions might include power generation, transmission, distribution, and load balancing. By mapping these functions, their relationships to each other, and the infrastructure that supports them, we can gain a clearer understanding of the system's vulnerabilities and potential points of failure.

Once the critical functions are identified, the next step is to associate each function with the resilience characteristics that most significantly contribute to its maintenance. This association is not necessarily one-to-one; multiple characteristics may support a single function, and vice versa. For example, the power distribution function in a grid might benefit from modularity (through segmented distribution networks), redundancy (via multiple transmission pathways), and plasticity (through adaptive load management). By explicitly linking these characteristics to specific functions, we can develop a more targeted approach to enhancing system resilience.

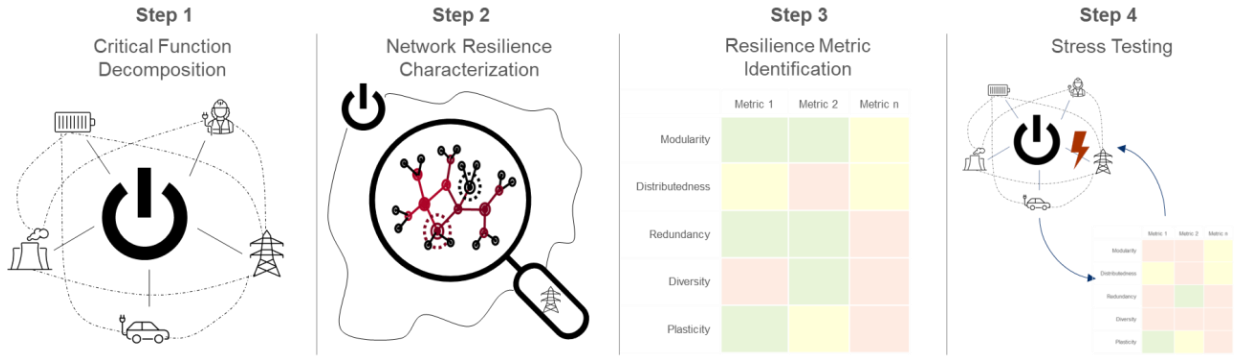

**Figure 2**. Placing threat-agnostic resilience characteristics within infrastructural critical functions through stress testing.

To quantify the degree to which each characteristic supports a given function, specific metrics need to be developed in the third step of the framework. These metrics should be measurable, reproducible, and sensitive to changes in system configuration. In the case of modularity in power distribution, a metric might include the number of independent distribution segments or the average size of network partitions. For redundancy, the metric could be the number of alternative transmission pathways between generation sources and end-users. By developing these quantitative measures, we can more accurately assess the resilience of the system and identify areas for improvement.

To evaluate the system's performance within and beyond these defined envelopes, stress testing and simulation techniques can be employed in step four. The system can be subjected to a wide range of stressors, not tied to specific threat scenarios, to evaluate its performance under diverse conditions. These stressors can include random node or link failures, resource constraints, or demand fluctuations. Advanced simulation techniques, such as agent-based modeling or Monte Carlo methods[49] can be used to explore the system's behavior under various conditions and identify potential vulnerabilities or failure modes.

This framework allows for a comprehensive assessment of system resilience without relying on predefined threat scenarios. By focusing on critical functions and their supporting characteristics, it provides a flexible approach that can adapt to emerging and unforeseen challenges. Moreover, it enables system designers and operators to identify key leverage points for enhancing resilience across multiple dimensions simultaneously. This approach is particularly valuable in the context of complex, interconnected infrastructure systems, where traditional risk-based approaches may be insufficient to capture the full range of potential disruptions and their cascading effects.

The intricate interplay of threat-agnostic resilience configurations within critical functions, as depicted in Figure 3, reveals a narrative of how complex infrastructure systems can mitigate losses, expedite recovery, and enhance adaptive capacity of the critical functions provided. Far from operating in isolation, these principles form a synergistic framework that amplifies the overall resilience of critical systems.

The reference system, depicted by a solid blue line, serves as a baseline for comparison, exhibiting a typical response pattern to disruption. This pattern is characterized by a sharp decline in performance following a disruptive event, marked by a yellow star, followed by a gradual recovery. Such behavior aligns with classical resilience models proposed by Holling[50] in 1973 and further developed by Walker et al.[39] in 2004. The reference system's trajectory enables a comparative analysis of systems enhanced with specific resilience characteristics, providing insights into the effectiveness of various resilience strategies.

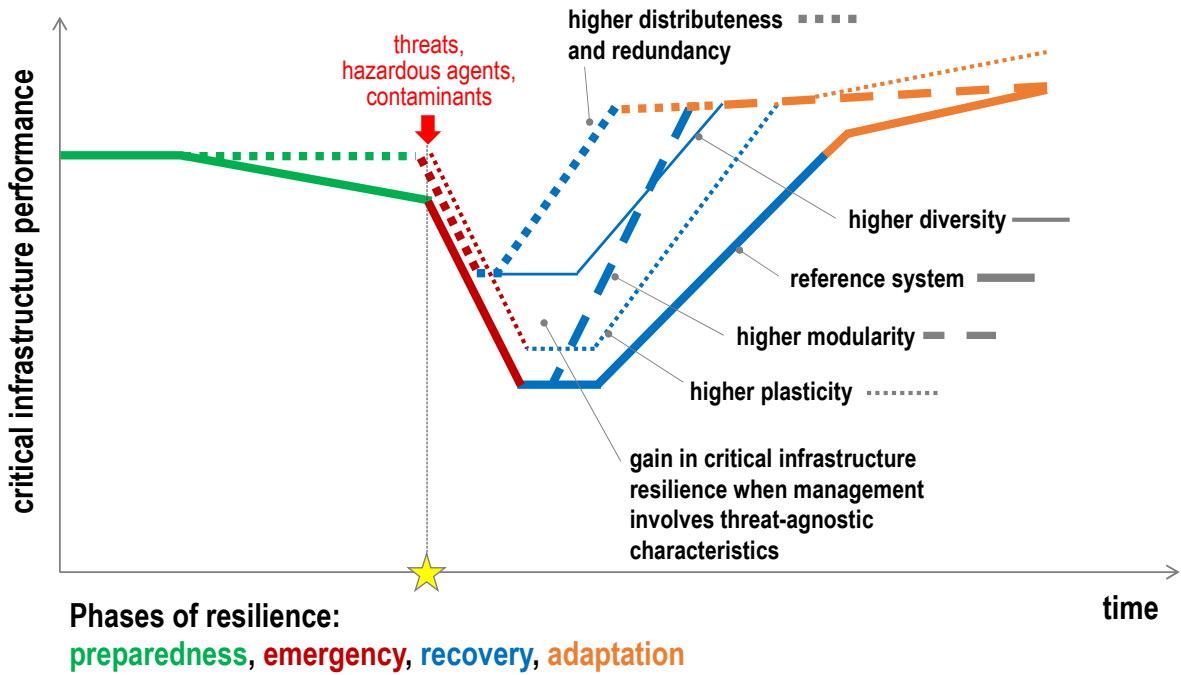

**Figure 3.** The positive impact of the five characteristics of threat-agnostic resilience on critical infrastructure performance

Higher modularity, represented by a dashed blue line, shows an intermediate response profile. The modular system experiences a less severe performance drop compared to the reference system and recovers at a moderate pace. Modular design facilitates infrastructure system mission execution amidst disruption, ranging from mitigating transit system delays during peak disruption periods[51], to rerouting of shipments in supply chains[52]. Likewise, the system with higher plasticity, depicted by a dotted blue line, exhibits a unique response profile characterized by a moderate initial performance decline but a rapid recovery. This behavior underscores plasticity's role in enabling quick system reconfiguration and adaptation to post-disruption conditions[53,54].

Systems with higher distributedness and redundancy, represented by a dotted blue line, demonstrate the most robust response to disruption. These systems experience a less severe initial performance drop and recover more rapidly, quickly surpassing the reference system's recovery trajectory. This superior performance can be attributed to the spatial dispersion of critical components and the availability of backup resources. Notable examples include municipal water systems, where centralization of piped water supply and sewer networks requiring central control are prone to systemic disruption from relatively minor disruptions to water quality that could often be addressed through local water treatment and management interventions[55]. The collective effect of these characteristics mitigates the impact of localized disruptions and accelerates the restoration process, highlighting the importance of decentralized design in critical infrastructure.

The system characterized by higher diversity, illustrated by a solid orange line, initially experiences a decline similar to the reference system with smaller losses and exhibits a steeper recovery curve. This behavior suggests that diverse systems, while not necessarily more resistant to initial shocks, possess a greater capacity for rapid adaptation and recovery. Emerging examples include municipal and regional energy grids, where systems with diverse energy sources recovered faster from major disruptions compared to homogeneous systems[56]. The varied resources and operational strategies inherent in diverse systems provide multiple pathways for recovery, enhancing overall system resilience[57].

.

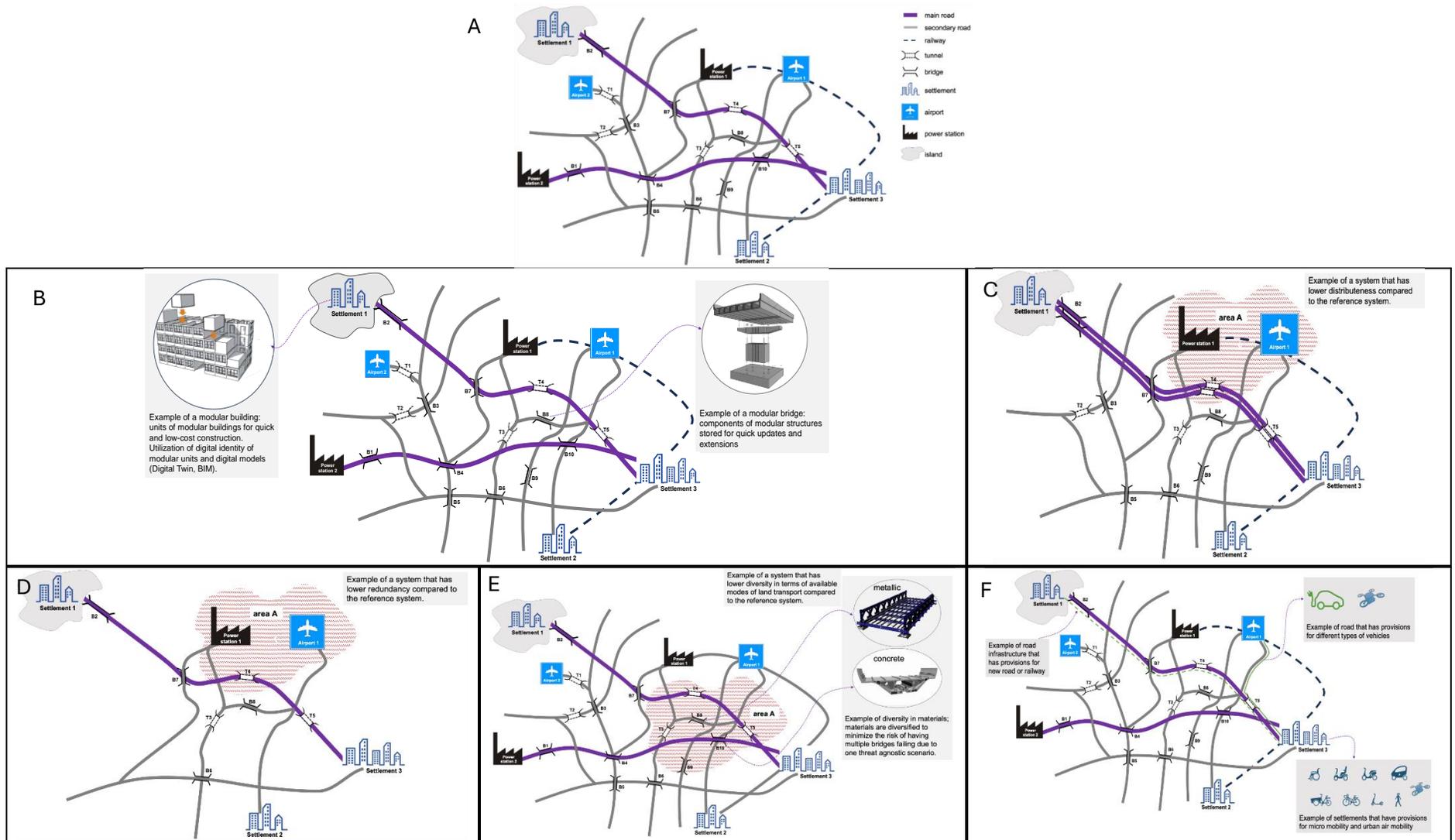

Figure 4. Threat-agnostic resilience within infrastructure. Panel A (top middle) presents a reference system of a transportation network connecting settlements, airports, and power stations. Panel B (center left), increases modularity by implementing modular construction materials and buildings. Panel C (center right) implements lower distributedness by removing a power station and airport. Panel D (bottom left) lowers redundancy by removing roadways, power stations, and airports. Panel E (bottom middle) increases diversity by varying construction materials based on local infrastructural use and bridge design, but limits other transportation modes to roadway. Panel F (bottom right) increases plasticity by adding new transportation modes.

Below, Figure 4 provides a comprehensive visual representation of how the principles of threat-agnostic resilience can be applied to complex, interdependent infrastructure systems. By illustrating various configurations of a critical infrastructure network, the figure demonstrates the impact of different resilience characteristics on system performance and adaptability in the face of unknown threats.

Panel A presents the reference system, which serves as a baseline for comparison. This configuration represents a typical infrastructure network, comprising multiple settlements, airports, power stations, and transportation links. The reference system exhibits a balanced approach to resilience, with a moderate level of redundancy, diversity, and distributedness. This baseline configuration allows us to evaluate how changes in system design can enhance or diminish overall resilience.

Panel B showcases a system with enhanced modularity compared to the reference system. In this configuration, we observe a more segmented structure, with clearly defined sub-systems that can operate independently, if needed. This design allows for localized management of resources and risks, ensuring that a disruption in one part of the system does not necessarily compromise the entire network. Second, the design facilitates easier maintenance and upgrades, as individual modules (units and components) can be taken offline for repairs or improvements without affecting the whole system (plug-and-play modules)[58].

Panel C illustrates a system with reduced distributedness compared to the reference system. In this scenario, critical functions and control centers are concentrated in fewer locations, presenting a stark contrast to the distributed approach of threat-agnostic resilience. While this centralized approach may offer some efficiency gains under normal operating conditions[59], it significantly compromises the system's resilience to unknown threats. The concentration of critical assets in Panel C creates a potential single point of failure, making the entire system vulnerable to localized disruptions. For instance, if area A (as indicated in the figure) is affected by an unforeseen event, the impact on the system could be far-reaching and limit future adaptive capabilities.

Panel D depicts a system with diminished redundancy compared to the reference case with area A illustrating a geospatial disruption concern for the system. In this configuration, we observe fewer backup components and alternative service routes. The system features only one airport and one power station, in contrast to the two of each, present in the reference system. The overall cost of the system is lower, in terms of construction and maintenance, however, its resilience is critically low. With fewer alternative paths and backup components, the system's ability to maintain functionality during disruptions is severely compromised. The lack of redundancy, specifically in area A, not only affects the system's ability to withstand disruptions but also impacts its recovery capacity. With fewer alternative resources available, the time and effort required to restore normal operations after a disruptive event would likely increase significantly.

Panel E focuses on the principle of diversity, showcasing a system with less varied modes of transportation compared to the reference system and more diverse construction materials. The lack of diverse transportation options in Panel E reduces the system's flexibility in responding to disruptions. For instance, if road networks are compromised due to an unforeseen event, the absence of alternative transportation modes could lead to significant isolation of certain settlements. Furthermore, Panel E hints at the importance of diversity at the component and material level. The example of constructing bridges using different materials, such as metallic and concrete, illustrates how diversity can enhance resilience against specific threats.

Panel F illustrates a system with greater plasticity compared to the reference case to accommodate better mobility through designated infrastructure. The connection between settlement 3 and airport 1 is enhanced with additional provisions to accommodate different types of mobility solutions. This flexibility allows the system to integrate new transportation technologies or adjust to changing travel patterns without requiring a complete overhaul of existing infrastructure. Similarly, the link between settlements 1 and 3 is designed with space and provisions for future expansion, such as the addition of a new road or railway. This foresight in planning enables the system to evolve organically in response to changing demands or technological advancements.

## Discussion and Conclusion

The analysis of threat-agnostic resilience characteristics and their application to critical infrastructure systems reveals several key insights with significant implications for infrastructure planning, design, and management. The rise of hybrid threats (e.g., socio-technical) calls for threat-agnostic approaches within infrastructure to prevent catastrophic failures[60,61]. Insights from network science provide the tools and methodologies to analyze and understand the structure, dynamics, and resilience of infrastructure[62-65]. By focusing on fundamental resilience characteristics through network science rather than specific threat scenarios, this approach offers a more comprehensive and flexible strategy for enhancing infrastructure performance across a wide range of potential disruptions which may not be identified by threat-aware assessments[66,67].

One of the key advantages of the threat-agnostic approach lies in its scalability and adaptability across diverse infrastructure sectors and geographical contexts. Whether applied to urban water systems, power grids, or transportation networks, the principles of threat-agnostic resilience provide a universal framework for improvement. This universality is particularly valuable for policymakers and infrastructure planners tasked with developing long-term strategies that can withstand evolving threats and changing societal needs. Moreover, the approach facilitates cross-sector collaboration and knowledge transfer, as resilience strategies developed for one infrastructure type can often be adapted and applied to others. By promoting a common language and set of principles for quantifiable and benchmarkable resilience, the threat-agnostic approach enables more effective coordination among different stakeholders involved in infrastructure development and management.

For infrastructure operators and managers, the threat-agnostic approach offers a more proactive stance on resilience. Rather than reactively addressing specific vulnerabilities as they are identified, this methodology encourages the continuous enhancement of system-wide resilience characteristics. This shift in focus can lead to more efficient resource allocation and a more holistic approach to risk management. By prioritizing system attributes, such as modularity and plasticity, operators can create infrastructure that is inherently more adaptable to changing conditions and emerging threats.

Equally, investors, insurers, and financial institutions stand to gain significant benefits from the adoption of a threat-agnostic approach to infrastructure resilience. By evaluating infrastructure projects through these lens, they can make more informed decisions about long-term viability and return on investment. Projects that demonstrate high levels of modularity, distributedness, redundancy, diversity, and plasticity may be viewed as more robust investments in an uncertain future. This perspective can lead to a shift in investment strategies, favoring projects that prioritize long-term resilience over short-term efficiency gains. Additionally, the threat-agnostic approach provides a more comprehensive framework for assessing and pricing risk in infrastructure investments, potentially leading to more accurate valuation of assets and more efficient allocation of capital in the infrastructure sector.

As data sensing infrastructure improves, the continuous monitoring and improvement of threat-agnostic resilience characteristics for long-term infrastructure health. The performance and condition of the system should be regularly assessed, using advanced sensing and data analytics technologies, to detect any potential vulnerabilities or inefficiencies[68]. The system should also be periodically updated and upgraded, incorporating new technologies and best practices, to keep pace with the evolving threats and opportunities in the environment and inform preparedness.

Governance and management of these resilience archetypes for infrastructure systems is subject for strategic and intervention-based decision-making. Pertinent decision-making processes should be agile and responsive, able to quickly detect and respond to the changing conditions, while balancing the trade-offs between the short-term and long-term objectives. The governance structures should also be adaptive and inclusive, engaging the diverse stakeholders and communities in the co-design and co-management of the systems, while ensuring the transparency, accountability, and fairness of the outcomes.

Despite its potential benefits, significant challenges remain in fully implementing the threat-agnostic approach to infrastructure resilience. Moving towards network principles requires substantial investments in the sensing, collection, integration and cleaning of data that is not universally available. This challenge is compounded by the complex, interdependent nature of modern infrastructure systems, which makes isolating and measuring the impact of individual resilience characteristics as well as their systemic corollaries difficult. However, investments towards threat agnostic resilience analysis are a necessity due to the exposure of complex infrastructure to ahistorical climatological and environmental stressors, a burgeoning global population, increasingly complex and interdependent economic activities, and the increasing disruptive potential for cyber and digital shock.

Implementing, tracking, and controlling threat-agnostic resilience within infrastructural systems requires deeper analysis based on the metrics for each resilience principle that this paper recommends. Governors, practitioners, and researchers alike may question the most favorable composition of any infrastructure system based on its setting. Moreover, balancing these characteristics together will require an individualized approach for any infrastructure system. Further research should uncover the steps necessary within stress-testing these resilience characteristics to determine the most practical, cost-effective resilience characteristics within individual and interconnected systems.

## Data availability

Datasets were not generated or analyzed in this article.

## Author contributions

**BT, SM, and SA fostered design concept and wrote the initial draft. GK, JP-O, RH, JT, and IL provided insight throughout initial draft and wrote revised text.**

## Competing interests

The authors have no competing interests to declare.